\documentclass[usenatbib]{mn2e}
\usepackage{graphicx}
\usepackage{subfigure}
\usepackage{rotating} 

\newcommand{\lsim }{{\lower0.8ex\hbox{$\buildrel <\over\sim$}}}
\newcommand{\gsim }{{\lower0.8ex\hbox{$\buildrel >\over\sim$}}}

\usepackage{soul}

\def\Chandra{\emph{Chandra}}

\def\RXTE{\emph{RXTE}}

\def\Swift{\emph{Swift}}

\def\XMM{{\it XMM-Newton}}

\def\j174540{CXO J174540.0-290005}

\def\simge{\mathrel{%
  \rlap{\raise 0.511ex \hbox{$>$}}{\lower 0.511ex \hbox{$\sim$}}}}
\def\simle{\mathrel{
  \rlap{\raise 0.511ex \hbox{$<$}}{\lower 0.511ex \hbox{$\sim$}}}}

\newcommand{\Msun}{\ifmmode {M_{\odot}}\else${M_{\odot}}$\fi}
\newcommand{\Lsun}{\ifmmode {L_{\odot}}\else${L_{\odot}}$\fi}
\newcommand{\Rsun}{\ifmmode {R_{\odot}}\else${R_{\odot}}$\fi}

\usepackage{xargs}
\usepackage{color}

\title[Very faint X-ray binaries]{The nature of very faint X-ray binaries; hints from light curves}
\author[Heinke et al.]{C.~O.~Heinke$^{1,2}$,  
A.~Bahramian$^1$, 
N.~Degenaar$^{3,4}$, 
R.~Wijnands$^5$ \\
$^{1}$ Dept. of Physics, University of Alberta, CCIS 4-183, Edmonton, AB T6G 2E1, Canada; heinke@ualberta.ca \\
$^2$ Alexander von Humboldt Fellow, at Max-Planck-Institut f\"ur Radioastronomie, Auf dem H\"ugel 69, D-53121 Bonn, Germany\\
$^3$ Department of Astronomy, University of Michigan, 500 Church Street, Ann Arbor, MI 48109, USA\\
$^4$ Hubble Fellow\\
$^5$ Anton Pannekoek Institute for Astronomy, University of Amsterdam, Postbus 94249, NL-1090 GE Amsterdam, The Netherlands\\
}

\begin{document}

\date{}

\pagerange{\pageref{firstpage}--\pageref{lastpage}} \pubyear{2014}

\maketitle

\label{firstpage}

\begin{abstract}
Very faint X-ray binaries (VFXBs), defined as having peak luminosities $L_X$ of $10^{34}$--$10^{36}$ erg/s, have been uncovered in significant numbers, but remain poorly understood. We analyse three published outburst lightcurves of two transient VFXBs using the exponential and linear decay formalism of King and Ritter (1998).  The decay timescales and brink luminosities suggest orbital periods of order 1 hour.  We review various estimates of VFXB properties, and compare these with suggested explanations of the nature of VFXBs. We suggest that: 
1) VFXB outbursts showing linear decays might be explained as partial drainings of the disc of ``normal'' X-ray transients, and many VFXB outbursts may belong to this category; 
2) VFXB outbursts showing exponential decays are best explained by old, short-period systems involving mass transfer from a low-mass white dwarf or brown dwarf; 
3) persistent (or quasi-persistent) VFXBs, which maintain an $L_X$ of $10^{34}$--$10^{35}$ erg/s for years, may be explained by magnetospheric choking of the accretion flow in a propeller effect, permitting a small portion of the flow to accrete onto the neutron star's surface. We thus predict that (quasi-)persistent VFXBs may also be transitional millisecond pulsars, turning on as millisecond radio pulsars when their $L_X$ drops below $10^{32}$ erg/s.

\end{abstract}

\begin{keywords}
accretion, accretion discs -- (stars:) binaries -- X-rays: binaries -- X-rays: individual: CXOGC J174540.0-290005 -- X-rays: individual: XMM J174457-2850.3
\end{keywords}

\section{Introduction}

Low-mass X-ray binaries (LMXBs) transfer matter from a low-mass star onto a neutron star or black hole, via an accretion disc.  Generally, those with the highest mass-transfer rates are persistent systems, while others are transients, spending most of their time in a quiescent state with little or no accretion, interrupted by occasional outbursts.  In persistent systems, relatively high mass transfer rates, and heating of the accretion disc through friction and X-ray irradiation, maintain the disc in an ionized state with high viscosity, permitting the mass to continue accreting at the rate it enters the disc \citep{Osaki74,White84,Lasota01,Coriat12}.  The outbursts of transient LMXBs often follow a fast-rise, exponential-decay shape \citep[e.g.][]{Chen97}, which can be understood in a disc instability model, in which continued accumulation of matter in the disc eventually ionizes the disc and raises its viscosity, leading to rapid dumping of the disc material \citep{Cannizzo82,Faulkner83,Meyer84,Huang89}.

Transient LMXBs in our Galaxy with relatively high peak X-ray luminosity $L_X$ ($\gsim4\times10^{36}$ ergs/s) are easy to find with a variety of all-sky monitors (e.g. the \RXTE\ ASM), if their X-rays are not too heavily absorbed by interstellar gas and dust.  A number of fainter transients, with peak $L_X$ of a few $\times10^{36}$ ergs/s, were detected by the Wide Field Cameras on {\it BeppoSAX}, in many cases showing thermonuclear bursts that prove the accretor is a neutron star \citep{Heise99}.  These systems showing fainter outbursts also tend to have shorter outbursts, with exponential folding times often $<$10 days \citep{intZand01,Cornelisse02a}.  It has been suggested that the majority of these systems contain partly degenerate (brown dwarf) donors, which have evolved past the orbital period minimum for hydrogen-rich donor stars \citep{King00}.

A fainter class of transients, with peak $L_X$ below $10^{36}$ ergs/s, came into clear view through observations with \Chandra, \XMM\ and \Swift\  \citep{Hands04,Muno05,Porquet05,Sakano05,Wijnands06,Degenaar09}.  These very-faint X-ray binaries (VFXBs) are mostly transients with quiescent luminosities below $10^{33}$ ergs/s, though some (apparently) persistent or quasi-persistent sources are known \citep[e.g.][]{intZand05,DelSanto07,Degenaar10b,ArmasPadilla13a}.  The VFXBs are likely connected to the ``burst-only'' sources identified through bursts with {\it BeppoSAX} without detectable ($>10^{36}$ ergs/s) non-burst emission \citep{Cocchi01,Cornelisse02a,Cornelisse02b,Campana09}. 

VFXBs are hard to understand, as they do not follow the standard patterns of behaviour in other X-ray binaries (as we show below), although X-ray binaries that usually show VFXB behaviour seem to be as numerous as X-ray binaries that typically show brighter outbursts \citep{Muno05}.  We note three ways\footnote{The behaviours of thermonuclear bursts at low accretion rates are also theoretically challenging \citep[e.g.][]{Peng07}, but we do not address that issue here.} in which VFXBs show unusual behaviours; 

i) When time-averaged accretion rates are measured (if multiple outbursts seen), they are often very low \citep{Degenaar09,Degenaar10}.  These low time-averaged accretion luminosities have been suggested to be very hard to explain in standard binary evolution models, which have difficulty reaching mass-transfer rates below $10^{-13}$ \Msun/yr within the age of the universe, as discussed in detail by \citet{King06}; but see section 3.1.

ii) They invariably have low peak X-ray luminosities, and low integrated luminosities over a single outburst. This indicates that the amount of material accreting from the disc at one time is small (see sections 2.3, 3.2).

iii) Persistent and quasi-persistent VFXBs maintain persistent mass transfer over long (years) periods of time, at rates far too low to maintain an irradiated disc. (This conclusion depends upon the orbital period, but see section 3.3).

 Several suggestions to explain the origin of VFXB behaviours have been proposed (mostly concentrating on explaining behaviour i).  In this paper, we analyze the lightcurves of some VFXB outbursts using the predictions for accretion disc behaviour from \citet{KingRitter98} (henceforth KR).
We then discuss the suggested explanations, and utilize the known information about these objects to improve our understanding of them. 

\section{Analysis of VFXB lightcurves}

\subsection{Formalism}
KR derive analytical expressions for the outburst lightcurve of a typical transient LMXB, predicting 
whether (for a given peak $L_X$ and outer disc radius) the lightcurve will follow an exponential or linear shape,
the timescale of the exponential decay $\tau_e$, 
the peak mass accretion rate (allowing calculation, for a given efficiency, of $L_X$), 
and the time when the exponential decay terminates, replaced with a linear decay that drops to (nearly) zero within another $\tau_e$.  
The details of these calculations were worked out in KR, made more rigorous by \citet{King98b}, extended to include evaporation by \citet{Dubus01}, applied to observations by \citet{Shahbaz98} and \citet{Powell07}, and supported by smoothed-particle-hydrodynamics accretion simulations in \citet{Truss02}.  (Alternative calculations have been worked out by, among others, \citealt{Lipunova00} and \citealt{Ertan02}, using different methods; we do not attempt to compare these here.) We summarize the key points here. 

In KR's disc model, the overall lightcurve shape is an exponential decline if irradiation by the central X-ray source is able to ionize the entire disc.   For a given outer disc radius $R_{11}$ (in units of $10^{11}$ cm), this produces critical luminosities above which the lightcurve should be exponential in shape  \citep[KR,][]{Shahbaz98}:
\begin{equation}
L_{crit}{\rm (NS)}=3.7\times10^{36} R_{11}^2 {\rm erg s}^{-1}
\end{equation}
and
\begin{equation}
L_{crit}{\rm (BH)}=1.7\times10^{37} R_{11}^2 {\rm erg s}^{-1}
\end{equation}
for neutron stars (NS) and black holes (BH) respectively.  The difference arises from whether the irradiation arises predominantly from a point source, applicable to accreting NSs; or a (foreshortened) inner disc (less efficient at irradiating the outer disc), applicable to BHs in the soft (disc-dominated) state.  Comparing the peak $L_X$ and lightcurve shape, thus, allows a constraint on the outer disc radius of an accreting system.

The peak luminosity of an outburst correlates with the amount of mass being transferred in the outburst, and thus with the size of the disc (at least, that participating in the outburst). Since LMXBs show different peak luminosities in different outbursts, it is likely that not all the accretion disc participates in all outbursts, and/or that outbursts are able to begin from significantly different disc masses \citep[e.g. KR,][]{Shahbaz98,Lasota01,Yu07,Degenaar10}. 
 For a given $R_0$($=R_{11}/10^{11}$ cm) participating in an outburst, the accretion rate at the beginning of an exponentially decaying outburst is initially estimated from the disc mass (assuming the critical density throughout) and viscous time (KR's equation 29):
\begin{equation}
\dot{M}= \phi R_{11}^{7/4} e^{-t/\tau_e} 
\end{equation}
where $\tau_e$ is the exponential decay time, $\tau_e=R_0^2/(3\nu)$, where $\nu$ is the viscosity, and $\phi$ encodes how much material is available for accretion in the disc, and how the disc is irradiated (see \citealt{Powell07}).  
\citet{Shahbaz98} showed that the total accreted mass was generally rather less (typically $\sim$10\%) than the theoretical maximum disc mass; in other words, $\phi$ should be lower than KR estimated.  
\citet{Powell07} found that choosing $\phi$ to be $1.3\times10^{-12}$ m$^2$ s J$^{-1}$ (=$1.3\times10^{-15}$ cm$^2$ s erg$^{-1}$)
produced reasonable matches to the lightcurves of most neutron star LMXB outbursts, though the observed black hole LMXB outbursts required $\phi$ values an order of magnitude lower. This might be explained by the less efficient illuminating geometry of soft-state black hole outbursts, or of other differences between black hole and neutron star systems. Below, we show that a small alteration to Powell et al's assumptions removes this discrepancy.

KR suggest $10^{15}$ cm$^2$ s$^{-1}$ for $\nu$.  This falls in the middle of the range of values derived by \citet{Shahbaz98} for $\nu$ from fitting observed LMXB lightcurves, which are typically lower for exponential decays, and higher for linear decays.
\citet{Powell07} select a value of $\nu=4\times10^{14}$ cm$^2$ s$^{-1}$, which places the inferred value of $R_0$ between the circularization radius $R_{circ}$ and the distance from the compact object to the inner Lagrange point, $b_1$, for the well-observed outbursts of SAX J1808.4-3658 and XTE J0929-314. This produces rough agreement between the circularization radius and inferred radius of the disc for all four of the relatively faint neutron star LMXB outbursts analyzed by \citet{Powell07}, though the bright neutron star and black hole outbursts seem to require $\nu$ to be a factor of $\sim$2 larger.  An increasing $\nu$ for larger-radius discs is reasonable, as $\nu$ is predicted to increase with $R$ (see, e.g., KR).

A small secondary maximum is sometimes observed in LMXB outbursts, and in this model may be attributed to a portion of the outer disc which is initially shielded from irradiation by, e.g., disc warps  \citep[KR,][]{Truss02}. This secondary maximum is expected to occur roughly $\tau_e$ after the peak; this can contaminate the fitting of the declines if the observing cadence is insufficient to resolve this (as is likely the case in some data we consider here).

KR show that after a few $\tau_e$ in outbursts showing exponential decay, the irradiation is insufficient to maintain the disc edge in an ionized condition, and the outburst shape changes to a faster decay, dropping into quiescence within roughly another $\tau_e$.  This transition produces a ``knee'' or ``brink'' feature in the lightcurve.  After this brink (or, if the entire disc is not irradiated, from the beginning of the outburst), the lightcurve follows a linearly decaying trend (with the possibility of additional minor outbursts, that again may muddy the fitting).  We fit the linear decay portion of the lightcurve with \citep{Powell07}
\begin{equation}
L(t) = L_t \left(1 - \frac{t-t_t}{\tau_l} \right)
\end{equation}
where $L_t$ is the luminosity at which the source transitions to a linear decay, $t_t$ is the time of the transition, and $\tau_l$ is the timescale of the linear decay.

\citet{Shahbaz98} studied outburst X-ray lightcurves of a number of transient LMXBs, and characterized them as showing exponential and/or linear decays.  Fitting them with the formalism mentioned above, Shahbaz et al. estimated physical parameters of the LMXB accretion discs, including the viscosity $\nu$ and the size and mass of the accreting discs.    These analyses were extended by \citet{Powell07}, specifically focusing on comparing the disc circularization radii (calculated from the known orbital parameters)
 and the inferred disc radii (using either the luminosity or exponential decay criterion).  Similar analyses have been done for Aquila X-1 \citep{Simon02,Campana13}, 4U 1608-52 \citep{Simon04}, transients in M31  \citep[e.g.][]{Williams04,Williams06}, the high-mass X-ray binaries V0332+53 \citep{Mowlavi06} and CI Cam \citep{Simon06}, and the accreting millisecond pulsars IGR J00291+5934 \citep{Torres08} and IGR J17511-3057 \citep{Falanga11}. These studies have tended to find general agreement, though the values of the viscosity $\nu$ and luminosity normalization $\phi$ appear to differ systematically for shorter vs. longer-period systems.  Nevertheless, the majority of transient LMXB outbursts have not been followed with the sensitivity and cadence required to resolve the predicted transition.

The interpretation of observed ``brinks'' as a transition from when the entire disc is ionized, to when it is not, is not the only potential explanation.  \citet{Powell07} identified a feature in SAX J1808.4-3658's lightcurve as this transition, while \citet{Shahbaz98} interpreted the same feature as a secondary maximum. \citet{Zhang98} and \citet{Campana98b} interpreted a brink in the lightcurve of Aquila X-1 as the transition from direct accretion onto the NS surface, to a ``propeller'' regime where the inner accretion disc is cut off at the magnetospheric radius \citep[see also, e.g.,][]{Hartman11,Asai13,Campana14}.  This interpretation allows the inference of the NS magnetic field, which in each case seems reasonable, so this interpretation cannot be ruled out. If the propeller interpretation of the brink is correct for any source, it means that the disc radius inferred from the brink should be considered as an upper limit for that source (since the brink when the outer edge stops being ionized would presumably happen at a lower luminosity).  We note, however, that brinks in the lightcurves of several black hole LMXBs (4U 1543-475, XTE J1550-564, GRO J1655-40, GX 339-4) have been suggested \citep{Powell07}, and brinks may also be visible in the lightcurves of, e.g., XTE J1908+094 \citep{Jonker04b} and XTE J1752-223 \citep{Russell12}. However, such brinks are not obvious in all X-ray transient lightcurves (and a variety of features could be produced by varying obscuration of the accretion disc, e.g. \citealt{Narayan05}), so this paradigm cannot be considered completely secure.


We will extend the analysis of Powell et al. to derive a minimum radius for two VFXBs with well-recorded transient outbursts.
When fitting the exponential decay, we use \citep{Powell07}:
\begin{equation}
L(t) = (L_t - L_e) \exp \left( -\frac{t-t_t}{\tau_e} \right) + L_e\\
\end{equation}
where $L_e$ is the limit of the exponential decay. We placed a simple constraint on the fitting, that $0.4 < L_e/L_t < 1.0$.

For our purposes (studying faint neutron star transients), we choose the value of the  parameter $\nu=4\times10^{14}$ cm$^2$ s$^{-1}$, as found by \citet{Powell07} to accurately describe the outbursts of faint neutron star transients.
From our adopted value of $\nu$, we compute 
\begin{equation}
R_0(\tau_e)=\sqrt{3\nu \tau_e}=3.5\times10^7 \sqrt{\tau_e}
\end{equation}

\begin{figure*}
\includegraphics[scale=0.35]{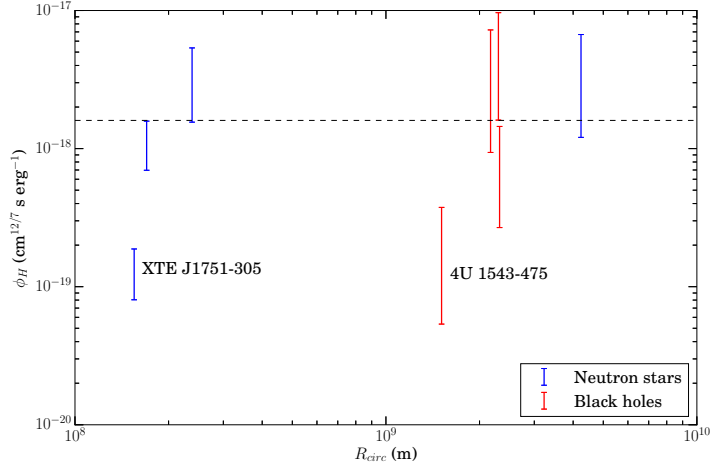}
\caption{Required values of $\phi_H$ to satisfy $R_{disc}^{12/7}=\phi_H L_t$, for eight transients with orbital information in \citet{Powell07}.  The ranges for each system reflect the $\phi_H$ value if $R_{disc}$ is either $R_{circ}$ (the lower $\phi_H$) or the primary Roche lobe radius $b$.  Other errors, such as the distance, are not included. The dashed horizontal line indicates our suggested value of $\phi_H$.}
\label{fig:phi}
\end{figure*}

For scaling the transition luminosity $L_t$ to the transition radius $R_{disc}$, Powell et al. implicitly assume that $R_{disc}^2=\phi L_t$ (their equation 21), which assumes that the disc thickness scales linearly with radius $R$.  This contradicts Powell et al's assumption (and that of KR) that the disc thickness $H$ scales as $R^{9/7}$. 
We instead use 
\begin{equation}
R_{disc}^{3-n}=R_{disc}^{12/7}=\phi_H L_t \\ 
\end{equation} 
(following Powell et al's equation 3), which is more physically consistent.  
As we have redefined $\phi_H$ (the subscript stresses this redefinition, accounting for the increasing thickness of the flaring disc), we recalibrate its value, based on the systems studied by Powell et al. (their Tables 2 and 3).  We find that choosing $\phi_H$ 
=$1.6\times10^{-18}$ cm$^{12/7}$ s erg$^{-1}$ is consistent ($R_0$ is between $R_{circ}$ and $b$) with the results from six of the eight systems with orbital information discussed in Powell et al. (see Figure 1). 
 For XTE J1751-305 and 4U 1543-475, smaller values of $\phi_H$, by a factor of 10, are required.  This may indicate that the brinks identified in these systems by Powell et al. are not due to this mechanism.
Powell et al. use the 1.5-12 keV $L_X$ for $L_t$, while we use 2-10 keV for the VFXBs; considering the scales of the other uncertainties (e.g. distance), this is not a major concern.

\subsection{Specific lightcurves}

The data for VFXB outbursts is significantly less complete than for brighter LMXB outbursts, due to the inability of typical all-sky monitors to detect VFXB outbursts, requiring pointed observations.  The available data is mostly from Swift/XRT monitoring observations of the Galactic Centre \citep{Degenaar09,Degenaar10}, as Swift/XRT provides high sensitivity and the capacity to undertake monitoring observations frequently.  We show here limited data on three outbursts by two VFXBs.  

The highest quality VFXB lightcurve is that of the 2013 outburst of CXO J174540.0-290005, reaching $L_X$(2-10 keV)$4\times10^{35}$ erg/s. The data were presented in \citet{Koch14}, including near-daily Swift/XRT observations along with multiple \Chandra\ and {\it NuSTAR} observations (see Figure 2).  We fit the 2013 outburst decline with an exponential, converting to a linear decline.  (We exclude two Swift datapoints around MJD=56457 from our fit; these appear to be a reflare.)  The exponential drop is relatively well-measured, and the linear decay timescale, though not well-constrained, is consistent.  For this outburst, we have enough data to simultaneously fit the exponential decay timescale $\tau_e$, the brink luminosity $L_t$ and its time $t_t$, and the timescale of the linear decay $\tau_l$ (see Table 1). The inferred radii are $1.4\pm0.2\times10^{10}$ cm (from the decay timescale) and $0.7^{+0.2}_{-0.1}\times10^{10}$ cm (from the brink luminosity), where the errors are propagated only from the fit.  Considering our lack of knowledge of $\nu$ and $\phi$, these are reasonably consistent.

\begin{figure*}
\includegraphics[scale=0.35]{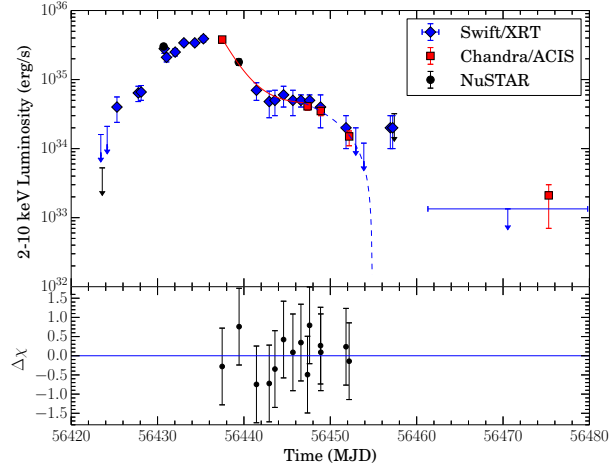}
\caption{Lightcurve of CXO J174540.0-290005's 2013 outburst (data from \citealt{Koch14}). The model fit includes an exponential decay (red solid line) and a linear decay (blue dashed line); see text for details. Residuals of the fit are plotted in the lower panel for the data used.}
\label{fig:174540_2013}
\end{figure*}

\begin{figure*}
\includegraphics[scale=0.35]{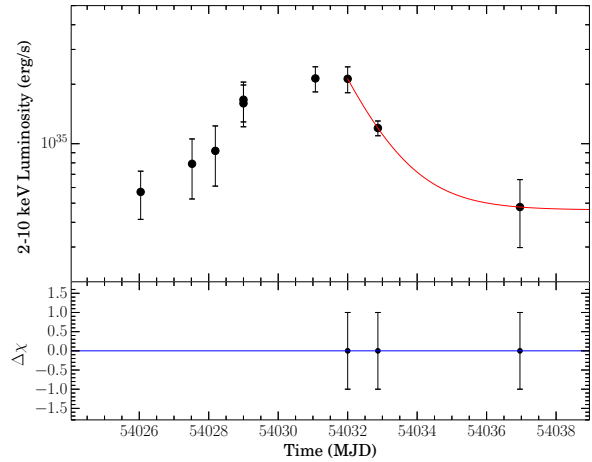}
\caption{Lightcurve of CXO J174540.0-290005's 2006 outburst (data from \citealt{Degenaar09}), fit with our model, as in Fig. 2.  Only the exponential decay is fitted. }
\label{fig:174540_2006}
\end{figure*}

The 2006 outburst of CXO J174540.0-290005 \citep{Degenaar09} was less frequently  sampled (Figure 3), with only three points defining the decay. The outburst appears to have a flat top with $L_X\sim2\times10^{35}$ ergs/s lasting $\sim$3 days (vs. $\sim$7 days for the brighter 2013 outburst), decaying to $\sim5\times10^{34}$ erg/s.  Since this $L_X$ range corresponds to the exponential part of the 2013 decay, we fit this decay with an exponential.  We are only able to effectively constrain the exponential decay timescale. We fix the limiting luminosity of the exponential decay to $4.8\times10^{34}$ ergs/s (assuming the same ratio of $L_t/L_e$ as in the 2013 outburst); changing this does not have a large impact.  
We find similar (possibly slightly smaller) values for the inferred radii as in the 2013 outburst, with larger uncertainties (see Table 1); below we use  $1.1\times10^{10}$ cm as our best estimate for CXO J174540.0-290005.

\begin{figure*}
\includegraphics[scale=0.35]{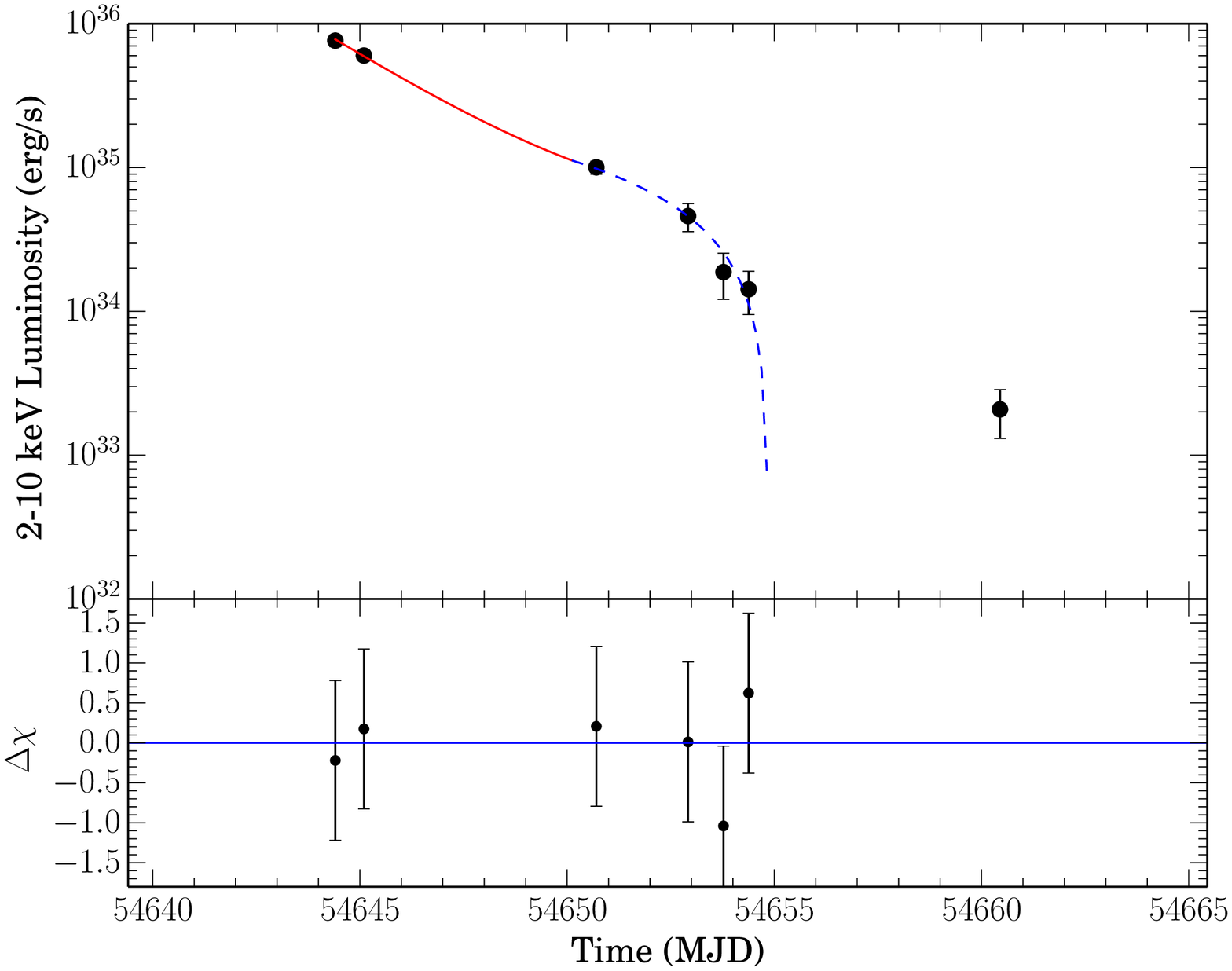}
\caption{Lightcurve of XMM J174457-2850.3's 2008 outburst (data from \citealt{Degenaar10}), fit with our exponential plus linear decay model, as in Fig. 2.}
\label{fig:174457_2008}
\end{figure*}

The 2008 outburst of XMM J174457-2850.3 \citep{Degenaar10} is the second most useful dataset, with a decay defined by six observations (Figure 4). We cannot be certain that we have observed the peak of the outburst, since the beginning of the outburst was not identified.  We can fit all the same quantities as in the 2013 outburst of CXO J174540.0-290005, though the transition from exponential to linear decay is not clearly resolved (Table 1).  The inferred radii are $1.6^{+0.1}_{-0.2}\times10^{10}$ cm (from the decay timescale), and $1.1^{+0.2}_{-0.4}\times10^{10}$ cm (from the brink luminosity); combining these, we estimate $\sim1.4\times10^{10}$ cm.  This transient is known to contain a neutron star, from the detection of a thermonuclear burst \citep{Degenaar14}.

The evidence for exponential decay in both systems argues that the entire disc was irradiated, and thus that we can estimate the true disc sizes from our fits.
 Assuming that these calculated radii correspond roughly to the circularization radii of the relevant discs (i.e., that the entire disc empties), and with a reasonable choice of the mass ratio $q\sim$0.1, we can then infer the orbital periods of these systems.
Rearranging equation 4.18 of \citet{Frank92}, we get 
\begin{equation}
P_{day}=(1/8) (R_{circ}/\Rsun)^{3/2} (1+q)^{-2} [0.500-0.227 \log q]^{-6} 
\end{equation}
For $q\sim$0.1, this simplifies to 
\begin{equation}
P_{day}=0.64* (R_{circ}/\Rsun)^{3/2} 
\end{equation}
Thus, we find predicted periods of  
 $\sim$1.0 hour for CXO J174540.0-290005, and $\sim$1.4 hours for XMM J174457-2850.3.

\subsection{Implications of Outburst Fluences, Peak Luminosities}

The low peak luminosities and integrated outburst luminosities can be used to infer quantitative conclusions.
We first address the integrated outburst luminosities of VFXBs, in general and for the transients studied above, using the formalism of \citet{King00}. \citet{King00} showed that the faint (peak $L_X < 10^{37}$ ergs/s) transients discovered by BeppoSAX in the Galactic Centre (e.g. SAX J1808.4-3658) were consistent with LMXBs that had evolved beyond their period minimum, agreeing with evolutionary modeling of specific sources  \citep[e.g.,][]{Bildsten01}.  We have not seen this calculation illustrated before for VFXBs.

The best-studied lightcurves of CXO J174540.0-290005 (2013 outburst) and XMM J174457-2850.3 (2008 outburst) are consistent with integrated fluences of $\sim5\times10^{-5}$ erg cm$^{-2}$ \citep{Koch14,Degenaar10}. For this fluence, a distance of 8 kpc, and a canonical (1.4 \Msun, 10 km) NS accretor\footnote{The accretor is certainly a NS for XMM J174457-2850.3 \citep{Degenaar14}; as many VFXBs are known to be NSs, this is a reasonable guess for CXO J174540.0-290005 as well.}, we can infer a total transferred mass of $2\times10^{-12}$ \Msun\ per outburst.   Using eq. 5 from \citet{King00}, we get  
\begin{equation}
M_{\rm transferred}=1.5\times10^{-11}m_1^{-0.35}\alpha_{0.05}^{-0.86} R_{10}^{3.05} \Msun\
\end{equation}
 where $m_1$ is the accretor's mass in units of \Msun, $\alpha_{0.05}$ is the disc viscosity normalized to 0.05, and $R_{10}$ is the disc radius in units of $10^{10}$ cm.  Again assuming a 1.4 \Msun\ NS accretor, we find 
\begin{equation}
R=5\times10^9 \alpha_{0.05}^{0.28} \ {\rm cm}
\end{equation}
Comparing this to the predicted size of the disc in an LMXB with a $\sim0.1$ \Msun\  companion (for example), 
\begin{equation}
R_{circ}=1.7\times10^{10} \left(\frac{P_{orb}}{80 {\rm min}}\right)^{2/3} \ {\rm cm} \\
\end{equation}
 we see that orbital periods of order 12 minutes are suggested, if the entire disc is accreted (as suggested by the exponential decays).  

Five Galactic Centre transients--the two objects studied above, plus Swift J174553.7-290347, Swift J174622.1-290634, and GRS 1741-2853 \citep{Degenaar09,Degenaar10}--showed outbursts with smaller integrated fluences, $<10^{-5}$ erg cm$^{-2}$.
In this case, we find 
\begin{equation}
R=3\times10^9 \alpha_{0.05}^{0.28} \ {\rm cm}
\end{equation}
 and infer orbital periods of order 6 minutes, if the entire disc is accreted.  However, three of these transients showed other outbursts with significantly larger fluences (GRS 1741-2853 showed one with peak $L_X \sim10^{37}$ ergs/s), indicating that the entire disc was not drained in these outbursts.  Indeed, only one of these outbursts (the 2006 outburst of CXO J174540.0-290005) shows any evidence for an exponential decay, and the evidence (only three data points) is weak, vulnerable to misinterpretation if a late rebrightening (or some other complexity in the lightcurve) occurred.

We can similarly compare a typical VFXB peak $L_X$ of $10^{35}$ ergs/s with relations between the peak $L_X$ and orbital period. KR present theoretical calculations of the mass transfer rate assuming X-ray irradiation of the disc, which leads to exponential decay.  Taking the mass ratio $q$=0.1, a disc radius of 0.7 times the tidal radius, and a 10 km, 1.4 \Msun\ NS, we can derive 
\begin{equation}
L_{\rm NS, peak, bol}=2.8\times10^{37} \left(\frac{P_{\rm orb}}{1 {\rm hr}}\right)^{7/6} {\rm ergs/s} \\
\end{equation}
  \footnote{We note that \citet{Maccarone13} claim that KR state $L_P=2.3\times10^{36} R_{11}$ ergs/s; this equation appears to be a misprint of  $L_P=2.3\times10^{36} R_{11}^2$, calculated in \citet{Wu10} to estimate the peak $L$ of systems which experience only a linear decay, based on equations 31 and 32 of KR.}
However, this calculation depends on the uncertain values of the disc density and viscosity, as well as assuming that the entire disc is ionized.   From KR's equation 31, we derive 
\begin{equation}
L_P=5.3\times10^{36} R_{11}^2=1.5\times10^{35}\left(\frac{P_{orb}}{1 {\rm hr}}\right)^{4/3} {\rm ergs/s} \\
\end{equation}
Allowing for a correction factor of 3 \citep{intZand07} for 2-10 keV $L_X$ to bolometric luminosity, we calculate a range of predicted orbital periods for a $L_{X,peak}=10^{35}$ ergs/s VFXB of 1 to 100 minutes. 

Alternatively, we can compare to empirical relations between the peak luminosity and orbital period. \citet{Wu10} found consistency of their data with a linear relation, which we rearrange to
\begin{equation}
\log P_{orb}=1.6\pm0.2 \log(L_{peak}/L_{Edd})+2.8\pm0.2 \\ 
\end{equation}
We thus estimate for CXO J174540.0-290005's 2013 outburst, assuming the factor of 3 correction to the 2-10 keV bandpass as above\footnote{Note that \citet{Wu10} used 3-200 keV X-ray luminosities.}, 
\begin{equation}
P_{orb}(L_{peak})=12^{+8}_{-5} \ {\rm minutes} \\
\end{equation}
 and for XMM J174457-2850.3, 34$^{+18}_{-11}$ minutes. 
An alternative relation with a saturated peak luminosity for orbital periods above 10 hours was suggested by \citet{PortegiesZwart04}, and updated by \citet{Wu10}; the updated version\footnote{No errors were provided for this relation, which was not found to be as good a description of the data as the linear fit.}  gives 
\begin{equation}
\log P_{orb}= 0.78 \log (L_{peak}/L_{Edd})+1.63 \\
\end{equation}
 This relation then predicts $P_{orb}(L_{peak})$=50 minutes for CXO J174540.0-290005, and 84 minutes for XMM J174457-2850.3, which are in relatively good agreement with our estimates from the lightcurve decay times. 
We caution that these estimates come from extrapolating an empirical relation outside the range in which it was validated. However, the low peak X-ray luminosities of these systems suggest, when compared to either theory or empirical trends, orbital periods below 2 hours. 



\section{Nature of VFXBs}

VFXBs may be an inhomogeneous group of objects. Thus, the unusual behaviours mentioned above may have different (sometimes overlapping) explanations.  We address these in turn below.

\subsection{Behaviour i): Low Time-Averaged Mass Transfer Rate}
 \citet{King06} specifically considered whether standard binary evolution--either ultracompact binaries (mass-losing white dwarfs), or cataclysmic variable-like late evolution (mass-losing main-sequence stars) could produce low time-averaged mass transfer rates, and concluded that with standard assumptions, rates of $<10^{-13}$ \Msun/year could not be reached within a Hubble time.

First, we check whether \citet{King06} used the correct mass transfer rate for comparison. Assuming a NS accretor (as verified for a number of VFXBs, see below), the inferred mass transfer rates from \citet{Degenaar10} (their Table 2) of the 8 transients in the Swift Galactic Centre monitoring region (all of which produced at least one VFXB outburst during the monitoring, i.e. $L_{peak}<10^{36}$ ergs/s) are: 
(3-8)$\times10^{-11}$, 
(7-14)$\times10^{-13}$, 
(2-6)$\times10^{-11}$, 
(1-10)$\times10^{-12}$, 
(5-20)$\times10^{-13}$, 
$5\times10^{-13}$, 
$\simle 1\times10^{-12}$, and 
$\simle 4\times10^{-13}$ \Msun/year.  
Of these, only two appear to be clearly below $10^{-12}$ \Msun/year, and none are required to be below $10^{-13}$ \Msun/year. (For CXO J174540.0-290005, \citet{Degenaar10} estimated (3-15)$\times10^{-13}$ \Msun/year. \citet{Koch14} used the presence of a second outburst, and a longer history, to better constrain its average mass transfer rate, quoting $7\times10^{-14}$ \Msun/year; we correct an error in their work to estimate $5\times10^{-13}$ \Msun/year.)

Accretion rates near $10^{-12}$ \Msun/year can be reached by standard binary evolution within a Hubble time.  One path is via ultracompact binary evolution, which reaches $10^{-12}$ \Msun/year in a couple Gyrs at orbital periods near 1 hour \citep{Iben85,Deloye03}, and can reach $10^{-13}$ \Msun/year in a Hubble time. The other path is cataclysmic variable-like post-period-minimum evolution, which produces roughly $\sim10^{-12}$ \Msun/year in 10 Gyrs at orbital periods near 2 hours \citep{Rappaport82,Howell01}.  \footnote{This estimate uses standard evolutionary sequences. It is not clear what the effects of additional sources of angular momentum loss \citep[e.g.][]{Patterson98}, or of inflated donor radii due to starspots \citep{Littlefair08}, may be.}  Since the orbital evolution in each case slows down with time, the majority of old X-ray binaries are expected to have mass-transfer rates below $10^{-11}$ \Msun/year \citep[e.g.][]{Stehle97}; this is consistent with the distribution of observed mass-transfer rates above.

For completeness, we consider the possibilities \citet{King06} suggested to explain the low time-averaged mass transfer rates of VFXBs;
 a) hydrogen-poor companions in old LMXBs containing black holes, 
b) low initial companion masses (e.g. brown dwarf companions when mass transfer starts), or 
c) intermediate-mass ($M\sim$1000 \Msun) black hole accretors.  
For many VFXBs, neutron star accretors have been verified through X-ray bursts (SAX J1828.5-1037, \citealt{Cornelisse02a}; 1RXS J171824.2-402934, \citealt{Kaptein00}; XMMU J174716.1-281048, \citealt{DelSanto07}; AX J1754.2-2754, \citealt{Sakano02,Chelovekov07}; 1RXH J173523.7-354013, \citealt{Degenaar10b}; IGR J17062-6143, \citealt{Degenaar13c}; and XMM J174457-2850.3, \citealt{Degenaar14}), ruling out that these objects are black holes, and thus that most VFXBs are black hole systems accreting from hydrogen-poor companions.
 The presence of VFXBs outside the centres of globular clusters (M15 X-3, \citealt{Heinke09b}; IGR 17361-4441, \citealt{Bozzo11}), while intermediate-mass black holes should rapidly sink to the cluster centres, also rules out intermediate-mass black holes for the known globular cluster VFXBs.  


\citet{Maccarone13} suggested that most VFXBs may be ``period-gap'' systems, in which the secondary has detached from its Roche lobe while evolving from a period of $\sim$3 down to $\sim$2 hours, and that mass transfer is due to a wind.  Accretion from the wind of main-sequence stars has previously been suggested for low-luminosity X-ray sources \citep{Bleach02,Pfahl02c,Willems03}. However, the wind-loss rates ($10^{-14}$-$10^{-16}$ \Msun/year) considered for M dwarfs (which compare reasonably to Proxima Cen's measured mass loss rate of $4\times10^{-15}$ \Msun/year, \citealt{Wood01}) are too low to explain the measured time-averaged mass transfer rates of the majority of known VFXBs, even before accounting for a wind accretion efficiency that is likely to be well below the Bondi-Hoyle-Littleton rate (see \citealt{Bleach02}).  

Declining radiative efficiency of advective flows at low luminosities, and thus in short-period black hole X-ray binaries, has been suggested repeatedly \citep[e.g.][]{Wu10,Maccarone13,Knevitt14}, and could help to explain the behaviour of some VFXBs suspected of containing black holes (e.g. Swift J1357.2-093313, \citealt{CorralSantana13}).  In Section 3.3 below, we discuss how reduced radiative efficiency (in the form of outflows, rather than advective flows) may also be a factor in some neutron star VFXBs.

\subsection{Behaviour ii): Outburst Fluences, Peak Luminosities}
The peak luminosities of VFXB transients suggest very small accretion discs, or that only part of the accretion discs are drained.  For some VFXB outbursts (e.g. the fainter outbursts of GRS 1741-2853 and AX J1745.6-2901), it is very likely that only part of the disc participated in the outburst, as discussed by \citet{Degenaar10} (cf. \citealt{Maitra08}).  Indeed, AX J1745.6-2901 has a known 8.4-hour eclipse period \citep{Maeda96,Porquet07}, proving the presence of a much larger disc.\footnote{This implies a high inclination, which might suggest that AX J1745.6-2901 was actually much brighter, and not really a VFXT. However, the detection of X-ray bursts reaching nearly the Eddington luminosity for a neutron star at the Galactic Centre \citep{Degenaar09,Degenaar12b} indicates that we see most of its $L_X$, and thus that it  does indeed show VFXB outbursts (its 2006 outburst, during which two X-ray bursts were detected, peaked at $9\times10^{35}$ ergs/s).  In any case, statistical arguments rule out that the majority of VFXBs are faint purely due to inclination effects \citep{Wijnands06}.}  \citet{ArmasPadilla13a} show that the optical/X-ray correlation from the likely black hole VFXB Swift J1357.2-0933 indicates that the disc is not strongly irradiated (but cf. \citealt{Shahbaz13}, who find evidence for an irradiated disc), suggesting that not all of the disc is heated, and thus supporting the suggestion that not all of the disc accretes in many VFXB outbursts. However, the exponential decays studied in \S 2 above indicate that, within the framework of the KR model, we can infer the disc sizes for some VFXB outbursts.

The calculations above based on peak luminosities suggest predicted orbital periods of 1 to 100 minutes.  
The integrated outburst luminosity calculations give similar predictions, of order 12 minutes for the particular systems studied here.
The empirical relations between peak luminosity and orbital period \citep{Wu10} predict between 12 and 84 minutes. 
However, the low range of these predictions ($<$30 minutes) gives an orbital period so small that the mass-transfer rate driven by (inescapable) gravitational radiation would be above $\sim10^{-10}$ \Msun/year.  For periods $<$20 minutes, the system could not even be transient in the disc instability model \citep[see ][]{Deloye03,Lasota08,Heinke13}).  

Our calculations of the exponential decay timescales and brink luminosities provide period estimates around 1 hour.  These are in excellent agreement with the evolutionary predictions of ultracompact evolution (discussed above), which give mass-transfer rates in the right range for this period range. There is even a perfect example of a borderline VFXB, the accreting millisecond pulsar NGC 6440 X-2, with a peak $L_X$ typically 1-2$\times10^{36}$ ergs/s, an orbital period of 57 minutes \citep{Altamirano10}, and a time-averaged mass transfer rate of $\sim10^{-12}$ \Msun/year \citep{Heinke10}.  

Alternatively, standard cataclysmic variable-like evolution gives periods not much longer ($\sim$2 hours) with mass transfer rates of $10^{-12}$ \Msun/year.  The poster child for such an evolution would be IGR J00291+5934, another accreting millisecond pulsar with peak $L_X$ values ranging from $2\times10^{36}$ down to $9\times10^{35}$ erg/s \citep{Hartman11}, an orbital period of 2.47 hours, a low-mass (partly degenerate) companion, and a time-averaged mass transfer rate of $\sim3\times10^{-12}$ \Msun/year \citep{Galloway05}.  We do not feel that the various indirect estimates of the orbital period discussed in this work are accurate enough to distinguish between a 1-hour orbit vs. a 2-hour orbit.  Spectroscopic (likely infrared) observations would be of great interest to search for the presence, or lack, of hydrogen lines, during outbursts from these systems.

\subsection{Behaviour iii): Persistent mass transfer at low rates}

\citet{intZand09} suggested that the persistent VFXB 1RXS J171824.2-402934 must be an ultracompact binary with orbital period $<$7 minutes, in order to maintain its disc in an ionized state at its average $L_X$ of $8\times10^{34}$ ergs/s. This would be consistent with the short orbital periods suggested by the low peak luminosities and integrated outburst luminosities of VFXB transients.  However, such short-period systems will have high mass-transfer rates due to gravitational radiation, and thus remain persistent, bright sources \citep{Deloye03,Lasota08}, leading to a clear contradiction.   
Furthermore, some persistent VFXBs show evidence of longer orbital periods;  
1RXH J173523.7-354013 (persistent at $2\times10^{35}$ ergs/s) 
shows strong H$\alpha$ emission in its optical spectrum \citep{Degenaar10b};
and M15 X-3 has an optical companion with a spectral energy distribution matching a 0.44 \Msun\ star \citep{Heinke09b,Arnason14}.
Somehow, the persistent VFXBs are continuing to accrete at very low rates.

\citet{Heinke09b} suggested that the propeller effect (the inhibition of accretion when the NS magnetosphere is rotating more quickly than the Keplerian orbital speed at the disc/magnetosphere boundary; \citealt{Illarionov75}) may be responsible for the inhibition of regular accretion in VFXBs.  Some current numerical work indicates that the propeller effect can build up a waiting (``trapped'') disc, rather than throwing material efficiently from the system \citep{D'Angelo10,D'Angelo12}.  Other numerical work indicates that the propeller effect may eject most of the infalling material from the binary \citep{Romanova03,Ustyugova06}.  In either case, a fraction (for instance, 10\% in some simulations of \citealt{Romanova04}) of the infalling material can still accrete continuously onto the star.  
Such a situation may explain the existence of persistent, or quasi-persistent, VFXBs. 

Observational evidence in favour of such a situation is the ``active'' quiescent state (with $L_X$ between $10^{33}$ and $10^{34}$ erg/s) of three ``transitional pulsars'', which have been seen as millisecond radio pulsars as well as in accreting states \citep{Archibald09,Papitto13,Patruno14,Bassa14,Bogdanov14}.  The rapid switches between this ``active'' quiescent state and a much lower-$L_X$ ``passive'' state \citep{Linares14} suggest that the X-ray luminosity in the ``active'' state is driven by a tongue of accretion down onto the NS, from a ``trapped'' disc.  \citet{Archibald14} recently identified pulsations during the ``active'' ($3\times10^{33}$ erg/s) state from the transitional pulsar PSR J1023+0038, which proves that accretion is continuing onto the NS surface.    \citet{Degenaar14} suggested that the VFXB XMM J174457-2850.3 (which we studied in \S 2 above) may also be a transitional pulsar, spending significant time in a similar ``active'' state around $10^{33}$--$10^{34}$ erg/s, in between periods of quiescence (at $L_X\sim5\times10^{32}$ erg/s) and outburst ($10^{35}$--$10^{36}$ erg/s).  

Here we suggest, for the first time, that many, or all, of the persistent and quasi-persistent VFXBs may be transitional pulsars in ``active'' states.  The transitional pulsar idea provides a reasonable explanation for how several quasi-persistent VFXBs could remain at $\sim10^{34}$ erg/s for years, turn ``off'' for one or more years, then resume their low luminosity.  
For instance, AX J1754.2-2754 appeared to be a persistent VFXB, but disappeared for up to a year before turning back on \citep{Bassa08,Jonker08}; it has also been suggested (on the basis of its low outburst optical magnitude) that this object should be ultracompact \citep{Bassa08}.  M15 X-3 was detected (retrospectively) by ROSAT in the mid-1990s at $10^{34}$ erg/s, then was faint ($\sim3\times10^{31}$ erg/s in 2001-2002, then returned to $10^{34}$ erg/s in seven observations between 2005 and 2013 \citep{Heinke09b,Arnason14}.  A possible problem with this interpretation is the lack of detection of pulsations in the persistent VFXB 1RXS J171824.2-402934 \citep{Patruno10}.

Some of the (quasi-)persistent VFXBs have time-averaged mass transfer rates that appear too low for the system parameters.  As an example, M15 X-3 has an estimated time-averaged mass-transfer rate of $2\times10^{-13}$ \Msun/year \citep{Heinke09b}, but the companion star is estimated to have a mass of $\sim$0.4 \Msun, allowing an inference of a roughly 4-hour orbital period \citep{Arnason14}.  
An active propeller could eject the majority of the inflowing material.  
Alternatively, irradiation-driven mass-transfer cycles, acting over periods of $\sim10^8$ years \citep[see e.g.][]{Hameury89,Buning04,Benvenuto14}, could alter the evolution of systems like M15 X-3.
These cycles are also a natural explanation for the substantially higher mass transfer rates in many persistent X-ray binaries, compared to those predicted by evolutionary models \citep{Podsiadlowski02}.  \citet{Ritter08} pointed out that irradiation-driven mass transfer instabilities only work if the system is persistent.  That is, when the system returns to its high accretion rate after a low state, it must be persistent, so as to sustain the irradiation over a thermal timescale; this does not exclude X-ray binaries from being transient during the low phase of their mass-transfer cycles.  

\section{Conclusions}

The nature of the objects producing VFXB outbursts (peaking between $\sim10^{34}$ and $10^{36}$ erg/s) is one of the key open questions in X-ray binary research, especially since these outbursts significantly outnumber ``normal'' ($>10^{36}$ erg/s) outbursts, and a majority of LMXBs in the Galactic Centre area have only been seen to exhibit VFXB outbursts. 
We applied the accretion disc lightcurve formalism of KR to the lightcurves of three VFXB outbursts, from two VFXBs in the Galactic Centre.  Particularly for the 2013 outburst of CXO J174540.0-290005, we found evidence for an exponential decay followed by a linear decay, in accord with the predictions of KR for a completely ionized disc. The timescale of the exponential decay, the luminosity of the ``brink'' where the linear decay begins, and the timescale of the linear decay, allow a rough inference of the accretion disc radius, and thus suggest an orbital period of order one hour for these two systems.  Most of the VFXB outburst lightcurves have too few points for reliable constraints, indicating the usefulness of daily Swift/XRT monitoring of the Galactic Centre, and daily monitoring of other VFXB outbursts after detection.

This inference depends on using the values for the viscosity, and luminosity scaling, parameters that \citet{Powell07} found accurately described the outbursts from several other short-period NS transients.  Estimates of the accretion disc size and orbital period from peak luminosities and total accreted mass give similar values, predicting orbital periods between 12 and 84 minutes.  Orbital periods below 30 minutes are inconsistent with the predicted high mass transfer rate of such short-period systems, within the disc instability outburst model.  

Comparing the time-averaged mass-transfer rates of well-studied VFXB transients with the predictions of evolutionary theory, we find that ultracompact binary evolution can easily reach the observed rates for orbital periods of 60-90 minutes, while CV-like evolution of hydrogen-rich companions evolving to larger periods after the $\sim$90-minute period minimum can also reach most of this range. 
Thus, we suggest that the most likely companions in transient VFXB systems are very low-mass white dwarfs, with orbital periods near 1 hour.  We also cannot rule out very low-mass brown dwarfs, with orbital periods near 2 hours, and recommend optical/infrared spectroscopy to search for H and He lines in VFXBs during outbursts.

A portion of VFXB outbursts are known to arise from LMXBs that also produce brighter outbursts. These are likely to involve only part of the accretion disc \citep{Degenaar10}. 
 A clear prediction of KR \citep[illustrated by]{Shahbaz98} is that such outbursts will show exclusively linear, not exponential, declines; it should be possible to distinguish linear vs. exponential decays in faint Galactic Centre transients with frequent (daily) monitoring.

The existence of persistent, and quasi-persistent, VFXBs is another challenge.  Keeping a disc ionized via persistent accretion at such low luminosities would require orbital periods of only a few minutes; but such short orbital periods would necessarily produce mass transfer rates orders of magnitude greater than observed, and two quasi-persistent VFXBs show definite evidence for hydrogen-rich donor stars.  We propose that a rapidly rotating NS magnetosphere chokes the accretion flow (producing a propeller, or a dead disc, situation), permitting only limited accretion.  This suggests that the quasi-persistent VFXBs may also be identified with transitional millisecond pulsars, turning on as radio pulsars when accretion is fully stopped and their X-ray luminosity drops below $10^{33}$ erg/s.
X-ray monitoring of these intriguing systems is thus suggested, to enable target-of-opportunity radio pulsation detections during low states.

\section*{Acknowledgments}
We thank the Swift team for their assistance in obtaining the observations that have made this work possible. 
COH is supported by an NSERC Discovery Grant, an Alberta Ingenuity New Faculty Award, and an Alexander von Humboldt Fellowship.  
ND is supported by NASA through Hubble Postdoctoral Fellowship grant number HST-HF-51287.01-A from STScI.  
We acknowledge extensive use of the ADS and arXiv.

\bibliographystyle{mn2e}
\bibliography{odd_src_ref_list}

\clearpage

\begin{table}
\begin{tabular}{lccl}
\hline
\hline
Source & \multicolumn{2}{l}{CXO J174540.0-290005} & XMM J174457-2850.3\\ 
Outburst & (2013)				& 	(2006)		& 2008 \\
\hline
$L_t$ ($10^{34}$ erg/s) & 4.6$^{+2.4*}_{-0.2*}$ & (5)$^@$ & 11$^{+3*}_{-6*}$ \\
$L_p$ ($10^{34}$ erg/s) & 39$\pm4$ & 21$\pm3$ & 76$\pm7$ \\
$t_t$ (MJD) & $56447^{+4}_{-5}$ & $54037^{+12}_{-1}$  & $54650^{+2*}_{-1*}$ \\ 
$\tau_e$ (days) & $1.8^{+0.7}_{-0.4}$ & $1.1^{+1.8}_{-0.5}$ & $2.4^{+0.1}_{-0.7*}$ \\
$\tau_l$ (days) & $8^{+8}_{-6*}$ & -$^a$ &  $4.7^{+0.4}_{-2.2*}$ \\
$R_0(\tau_e)$ & $1.4\pm0.2\times10^{10}$ & $1.1^{+0.7}_{-0.3}\times10^{10}$  & $1.6^{+0.1}_{-0.2}\times10^{10}$  \\  
$R_0(L_t)$ & $0.7^{+0.2}_{-0.1}\times10^{10}$  &  - &  $1.1^{+0.2}_{-0.4}\times10^{10}$ \\
$P (\tau_e)$ & $1.4^{+0.3}_{-0.3}$ & $1.0^{+1.0}_{-0.4}$ & $1.7^{+0.2}_{-0.3}$ \\
$P (L_t)$ & $0.5^{+0.2}_{-0.1}$ & - & $1.0^{+0.2}_{-0.5}$ \\
$P (L_{peak,1})$ & $0.2^{+0.1}_{-0.1}$ & - & $0.6^{+0.3}_{-0.2}$\\
$P (L_{peak,2})$ & 0.8 & - & 1.4 \\
\hline
\end{tabular}
\caption{Parameters of our fits to the lightcurves of three VFXT outbursts.  $L_p$: peak luminosity (2-10 keV) of outburst. $L_t$: brink luminosity (change from exponential to linear decline).  $\tau_e$: timescale of exponential decline. $\tau_l$: timescale of linear decay.  $R_0(\tau_e)$: accretion disc radius inferred from exponential decay timescale (see text), in cm.  $R_0(L_t)$: accretion disc radius inferred from brink luminosity (see text), in cm.   Errors are 90$\%$ uncertainties. $P$ are periods, in hours, estimated using the methods listed; $L_{peak,1}$ is the linear relation from \citet{Wu10}, while $L_{peak,2}$ is the linear relation with saturated peak luminosity for $P_{orb}$ above 10 hours, also from \citet{Wu10}.
* represents hard limits (limits reached by model constraints and not actual 90$\%$ uncertainty). $^@$ indicates the brink was not observed in this case.
}
\end{table}

\end{document}